\documentstyle[11pt,epsfig]{article}

\input{tcilatex}
\begin{document}

\begin{titlepage}
\begin{center}
\vspace{1cm}
\hfill
\vbox{
    \halign{#\hfil         \cr
      IPM/P-2003/042\cr
      hep-th/0308050\cr
      Aug, 2003\cr} }
      
\vskip 1cm
{\Large \bf
Scalar Solitons in Non(anti)commutative Superspace  }
\vskip 0.5cm
{\bf Reza Abbaspur\footnote{e-mail:abbaspur@theory.ipm.ac.ir} 
}\\ 
\vskip .25in
{\em
Institute for Studies in Theoretical Physics and Mathematics,  \\
P.O. Box 19395-5531,  Tehran,  Iran.\\}
\end{center}
\vskip 0.5cm

\begin{abstract}
We study solitonic solutions of a deformed Wess-Zumino model  in 2 dimensions, 
corresponding to a deformation of  the usual ${\cal N}=1,D=2$ superspace to the one with 
non-anticommuting odd supercoordinates.The deformation turns out to add a kinetic term for 
the auxiliary field besides  the known $F^{3}$ term coming from the deformation of the 
cubic superpotential.  Both these modifications are proportional to the effective deformation
parameter $\lambda \equiv \det C$, where $C$ denotes the non-anticommutativity matrix.
We find a modified ``orbit'' equation which on the EOM relates the
auxiliary and the scalar components of the scalar superfield as a first order correction 
to the usual relation in terms of the small parameter $\lambda $. Subsequently, we obtain 
the modified form of the first order BPS equation for the scalar field and find its solution to 
first order in $\lambda $. Issues such as modification of the BPS mass formula and a
non-linear realization of the ${\cal N}=1$ supersymmetry are discussed.

\end{abstract}

\end{titlepage}

\section{Introduction}

There have been several occasions in string theory where the study of string
dynamics near certain corners of the moduli space of the theory has led to
some new insights in field theory. One of the most known examples is the
emergence of spacetime noncommutative field theories from string theory in
the presence of an NS-NS background field \cite{1} (see e.g. \cite{2} for a
review). Another example has been provided more recently by the work of
Seiberg \cite{3} (and also \cite{4}), inspired by earlier works of Ooguri
and Vafa \cite{5,6} in the context of the gauge theory/ matrix model
correspondence (see e.g. \cite{7}-\cite{9}). Both these works (as well as
some others such as \cite{10}) have derived a new type of
non(anti)commutative field theories from string theory in the presence of a
constant graviphoton background field. These can be formulated as field
theories on a superspace whose grassmann odd supercoordinates obey a
Clifford like algebra, rather than the usual (anti)commutation relations.
This type of noncommutativity had also been considered earlier mainly from
an algebraic point of view \cite{11}-\cite{13}. This is an alternative
deformation in contrast to the ordinary noncommutativity among the bosonic
coordinates of superspace \cite{13a}-\cite{17}.

The work by Seiberg has initiated a new trend of research on various aspects
of field theories formulated on such deformed superspaces, which usually (in
a 4-dimensional context) are known as ${\cal N}=1/2$ theories \cite{3,4}
(for a partial list see \cite{18}-\cite{26}). In particular, the
deformations of some ${\cal N}=1$ theories (with ${\cal N}=\frac{1}{2}$
SUSY), such as the WZ model and SYM theories with or without matter have
been investigated in \cite{3,18,19}. The deformation of ${\cal N}=2$ SYM\
theories (with ${\cal N}=\frac{1}{2}+\frac{3}{2}$ SUSY) has also been
considered in \cite{4,20}. Some perturbative aspects of this class of
theories, such as their renormalizability, has been recently studied in a
number of papers \cite{21}-\cite{25}. Most of these papers deal with
formulation of deformed theories in a euclidean superspace. The Minkowskian
case has been considered in ref.\cite{26}. A combination of the
noncommutativity among both the bosonic and fermionc coordinates of
superspace has also been studied in \cite{11,22,22a}.

While by now a rather rich collection of information on perturbative aspects
of the deformed theories on non(anti)commutative superspaces has been
obtained, yet less has been known about their non-perturbative aspects such
as their solitonic solutions of the form of lumps, monopoles, instantons,
etc. (see however the comments in \cite{3}). It is interesting, for example,
to know to what extent such solutions parallel their counterparts in the
ordinary noncommutative field theories \cite{27} (for a review of the latter
see, e.g., \cite{28}). As a first step in this direction, in this paper we
explicitly work out the example of 1/2 supersymmetric BPS solutions in a
deformed Wess-Zumino model in 2 dimensions, whose commutative counterpart
has been studied long ago by Witten and Olive \cite{29}$.$\footnote{%
We will work throughout this paper with a euclideanized version of the WZ
model in \cite{29} in order to adapt the notation we had used in the
previous works \cite{14}-\cite{17}.} In 2 dimensional deformed WZ model, as
in its 4-dimensional counterpart, a modification proportional to $F^{3}$
(with $F$ as the auxiliary component of the scalar superfield) arises which
breaks all the ${\cal N}=1$ supercharges explicitly. But here, unlike in
4-dimensional case, also a kinetic part for $F$ appears which makes it a
dynamical field in contrast to the undeformed theory and it also breaks the
whole supersymmetry explicitly.

The reason why we adhere to 2 spacetime dimensions is that the deformation
of superspace by non-anticommuting odd coordinates in a theory in any
dimensions does {\it not} violate the conditions of Derrick theorem \cite{30}%
, as it does not bring higher order derivatives into the spacetime
formulation of the theory. As a result, scalar solitons (or lumps) can
indeed do exist {\it only} in 2 spacetime dimensions just as the case in the
undeformed theory. This is in contrast to the ordinary noncommutative
solitons where the violation of Derrick theorem due to higher order
derivatives allows for the existence of scalar solitons in higher spacetime
dimensions \cite{27}.

As we mentioned above, in the 2-dimensional case, the deformation breaks
both the two supersymmetries of the theory as a result of the fact that
2-dimensional SUSY generators do {\it not} anti-commute with the grassmann
odd derivatives appearing in the definition of the deforming star product
algebra. In the same spirit of \cite{3,4}, we may like to call the deformed
model as an ${\cal N}=0$ or non-supersymmetric theory. However, we shall
find that, on a certain surface relating the auxiliary and scalar fields
(called as the ``orbit''), the model still preserves the ${\cal N}=1$ SUSY
somehow non-linearly. This should be compared to a non-linear realization of
the broken ${\cal N}=\frac{3}{2}$ SUSY in the case of the 4-dimensional
deformed ${\cal N}=2$ SYM\ theory found in \cite{4}.

The organization of this paper is a follows: In sec.2 we review some
elementary facts about the deformed algebra of functions of two grassmann
odd variables. We then introduce in sec.3 the deformed generators of the $%
{\cal N}=1$ SUSY and the corresponding superderivatives on the basis of
which we construct the deformed WZ model. In sec.4 we obtain the modified
BPS equation governing continuous deformation of the ordinary BPS solutions
in the undeformed WZ\ model and solve them to first order in the deformation
parameter. In sec.5 we obtain an expressions for a classical effective
superpotential arising in the deformed WZ model, using which we compute the
deformation of the usual BPS mass formula. Finally, in sec.6 we show that
the ${\cal N}=1$ SUSY has a non-linear realization in terms of an effective
non-linear $\sigma $-model describing the dynamics on a relevant orbit in
the deformed theory.

\section{The deformed Grassmann Algebra}

In this section we review some basic facts regarding the deformed algebra of
functions of two grassmann odd variables $\theta ^{\alpha }\equiv (\theta
^{+},\theta ^{-})$ defining a $SO(2)$ spinor, corresponding to the algebra
of superfields on a deformed ${\cal N}=1,$ $D=2$ euclidean superspace (we
will use notations of. \cite{16,17}). The starting point for defining the
deformed algebra is the non(anti)commutation relation between the odd
variables, written as the Clifford algebra 
\begin{equation}
\{\hat{\theta}^{\alpha },\hat{\theta}^{\beta }\}=C^{\alpha \beta },
\label{1}
\end{equation}
in which $C^{\alpha \beta }=C^{\beta \alpha }$ is a constant bispinor and
the hats are used to interpret $\theta $'s here as operators. Defining the
algebra of functions of odd coordinates via the odd star product as \cite{3} 
\begin{equation}
f(\theta )*g(\theta )\equiv f(\theta )\exp \left( -\frac{1}{2}C^{\alpha
\beta }\overleftarrow{\frac{\partial }{\partial \theta ^{\alpha }}}\,%
\overrightarrow{\frac{\partial }{\partial \theta ^{\beta }}}\right) g(\theta
),  \label{2}
\end{equation}
one then finds a representation of the Clifford algebra (\ref{1}) on the
space of ordinary (anticommuting) grassmann odd variables. Here the left and
right differentiations on a function $f(\theta )$ of even or odd grassmann
party are simply related as 
\begin{equation}
f(\theta )\overleftarrow{\frac{\partial }{\partial \theta ^{\alpha }}}%
\,=(-1)^{\deg f}\overrightarrow{\frac{\partial }{\partial \theta ^{\beta }}}%
f(\theta )  \label{2a}
\end{equation}
where $\deg f=0,1({\rm {mod}\;2)}$ for $f$ even or odd, respectively. We
have to stress that distinguishing between the left and right
differentiations in eq.(\ref{2}) is important in order to have explicit
isomorphism between the algebra (\ref{2}) and the Clifford algebra and in
proving the associativity of the star product. In particular, it can be
easily verified that 
\begin{eqnarray}
\theta ^{\alpha }*\theta ^{\beta } &=&\theta ^{\alpha }\theta ^{\beta }+%
\frac{1}{2}C^{\alpha \beta },  \label{3} \\
(\theta ^{\alpha }*\theta ^{\beta })*\theta ^{\gamma } &=&\theta ^{\alpha
}*(\theta ^{\beta }*\theta ^{\gamma })=\theta ^{\alpha }\theta ^{\beta
}\theta ^{\gamma }+\frac{1}{2}(C^{\alpha \beta }\theta ^{\gamma }+C^{\beta
\gamma }\theta ^{\alpha }-C^{\gamma \alpha }\theta ^{\beta }).  \nonumber
\end{eqnarray}
(Note that in our 2d case the term $\theta ^{\alpha }\theta ^{\beta }\theta
^{\gamma }$ in the second relation identically vanishes.) Using the ordinary
rules of the grassmann calculus one can then see that the above defining
equation for the star product terminates at a finite order of $\partial
/\partial \theta $ 's. In particular, the exponential expansion in this
equation takes the simple form 
\begin{equation}
f(\theta )*g(\theta )=fg+(-1)^{\deg f}\frac{1}{2}C^{\alpha \beta }\frac{%
\partial f}{\partial \theta ^{\alpha }}\frac{\partial g}{\partial \theta
^{\beta }}-\frac{1}{4}\det C\frac{\partial ^{2}f}{\partial \theta ^{2}}\frac{%
\partial ^{2}g}{\partial \theta ^{2}},  \label{4}
\end{equation}
where now a grassmannian derivative is understood to act from the left (i.e. 
$\frac{\partial }{\partial \theta ^{\alpha }}=\overrightarrow{\frac{\partial 
}{\partial \theta ^{\alpha }}}$) and we have introduced the notation $\frac{%
\partial ^{2}}{\partial \theta ^{2}}\equiv \frac{\partial ^{2}}{\partial
\theta ^{+}\partial \theta ^{-}}\equiv \frac{1}{2}\frac{\partial }{\partial
\theta ^{\alpha }}\frac{\partial }{\partial \theta _{\alpha }}$ (with $%
\theta _{\alpha }\equiv \epsilon _{\alpha \beta }\theta ^{\beta }$)$.$

Also, similar to the case of the ordinary (bosonic) star product \cite{1,2},
one can show that the ``inner product'' of two functions of $\theta $ is
independent of $C^{\alpha \beta }$, namely 
\begin{equation}
\int d^{2}\theta \;f(\theta )*g(\theta )=\int d^{2}\theta \;f(\theta
)g(\theta ),  \label{5}
\end{equation}
which corresponds to the fact that the difference 
\begin{equation}
f(\theta )*g(\theta )-f(\theta )g(\theta )=\frac{\partial }{\partial \theta
^{\alpha }}(\cdot \cdot \cdot )^{\alpha }  \label{6}
\end{equation}
is a total grassmannian derivative not surviving the grassmannian
integrations. Often we denote this equivalence relation symbolically as $%
f*g\cong fg$.

We can find explicit expressions for powers of $f(\theta )$ under the star
product using the general expression given in eq.(\ref{4}). For this, we
first recall the basic expression 
\begin{equation}
f_{*}^{2}=f*f=f^{2}-\frac{1}{4}\det C\left( \frac{\partial ^{2}f}{\partial
\theta ^{2}}\right) ^{2},  \label{7}
\end{equation}
from which, for example, the third power of $f$ follows as 
\begin{equation}
f_{*}^{3}=f*(f_{*}^{2})=f^{2}-\frac{1}{4}\det C\left[ 3f\left( \frac{%
\partial ^{2}f}{\partial \theta ^{2}}\right) ^{2}+2\frac{\partial f}{%
\partial \theta ^{+}}\frac{\partial f}{\partial \theta ^{-}}\frac{\partial
^{2}f}{\partial \theta ^{2}}\right] ,  \label{8}
\end{equation}
and so forth for $f_{*}^{n}$. We note that all such expressions can totally
be written in terms of the second derivatives of $f$ and $f^{2}$ using the
identity 
\begin{equation}
\frac{\partial ^{2}f^{2}}{\partial \theta ^{2}}=2\frac{\partial f}{\partial
\theta ^{+}}\frac{\partial f}{\partial \theta ^{-}}+2f\frac{\partial ^{2}f}{%
\partial \theta ^{2}}.  \label{9}
\end{equation}
In this 2-dimensional case, further simplifications occur in computing
higher powers of $f$ (or generally any function of $f$) due to the fact that 
$\frac{\partial ^{2}}{\partial \theta ^{2}}$ of every function of $\theta $
is independent of $\theta $. A generic feature of such expressions for $%
f_{*}^{n}$ is that they depend on $C^{\alpha \beta }$ only through its
determinant $\det C=\frac{1}{2}C^{\alpha \beta }C_{\alpha \beta }$, which is
a Lorentz invariant quantity. Also all such expressions contain a total
derivative part which can be neglected upon grassmannian integrations. Often
we can separate the total derivative part of $f_{*}^{n}$ using the lemma $%
\int d^{2}\theta \;f*g=\int d^{2}\theta \;fg$ and the above expression for $%
f_{*}^{2}$. For example, using 
\begin{eqnarray}
\int d^{2}\theta \;f_{*}^{2} &=&\int d^{2}\theta \;f^{2},  \nonumber \\
\int d^{2}\theta \;f_{*}^{3} &=&\int d^{2}\theta \;ff_{*}^{2}=\int
d^{2}\theta \;f\left[ f^{2}-\frac{1}{4}\det C\left( \frac{\partial ^{2}f}{%
\partial \theta ^{2}}\right) ^{2}\right] ,  \nonumber \\
\int d^{2}\theta \;f_{*}^{4} &=&\int d^{2}\theta \;(f_{*}^{2})^{2}=\int
d^{2}\theta \;\left[ f^{2}-\frac{1}{4}\det C\left( \frac{\partial ^{2}f}{%
\partial \theta ^{2}}\right) ^{2}\right] ^{2},  \label{10}
\end{eqnarray}
we conclude the identities 
\begin{eqnarray}
f_{*}^{2} &\cong &f^{2},  \nonumber \\
f_{*}^{3} &\cong &f^{3}-\frac{1}{4}\det C\,f\left( \frac{\partial ^{2}f}{%
\partial \theta ^{2}}\right) ^{2},  \nonumber \\
f_{*}^{4} &\cong &f^{4}-\frac{1}{2}\det C\,f^{2}\left( \frac{\partial ^{2}f}{%
\partial \theta ^{2}}\right) ^{2}+\frac{1}{16}(\det C)^{2}\left( \frac{%
\partial ^{2}f}{\partial \theta ^{2}}\right) ^{4}.  \label{11}
\end{eqnarray}
As these expressions suggest, ignoring total derivative terms, the deformed
version of $f^{n}$ (as well as for any analytic function of $f$) is always
written as a polynomial (generally a power series) in powers of $\det C$. It
is indeed this beauty of the star product in reducing all the $C$-dependence
to the Lorentz invariant combination $\det C$ which causes a great deal of
simplicity in formuation of a deformed Wess-Zumino model in 2-dimensions in
later sections. This is also consistent with the fact that the deformation
defined by $C^{\alpha \beta }$ does not break Lorentz invariance of the low
energy field theory \cite{3,5}.

\section{The deformed Supersymmetry in Noncommutative ${\cal N}=1,$ $D=2$
Superspace}

Consider first the ordinary SUSY transformations acting on a superfield $%
S(x,\theta )$ via the generators\footnote{%
In the notation of \cite{16}, \cite{17} the matrices $\gamma ^{\mu }$ in
terms of the 2-dimensional Dirac matrices are written as $\gamma ^{\mu
}=\rho ^{1}\rho ^{\mu }$.} 
\begin{equation}
{\cal Q}_{\alpha }=\frac{\partial }{\partial \theta ^{\alpha }}+i\gamma
_{\alpha \beta }^{\mu }\theta ^{\beta }\partial _{\mu },  \label{12}
\end{equation}
which in components are written as 
\begin{equation}
{\cal Q}_{\pm }=\frac{\partial }{\partial \theta ^{\pm }}+i\theta ^{\pm
}\partial _{\pm }.  \label{13}
\end{equation}
The deformed SUSY is defined as the transformations generated by acting $%
{\cal Q}$'s on $S(x,\theta )$ via the star product operation, namely by 
\begin{eqnarray}
{\cal Q}_{\alpha }*S &=&\left( \frac{\partial }{\partial \theta ^{\alpha }}%
+i\gamma _{\alpha \beta }^{\mu }\theta ^{\beta }\partial _{\mu }\right) *S 
\nonumber \\
&=&\frac{\partial S}{\partial \theta ^{\alpha }}+i\gamma _{\alpha \beta
}^{\mu }\partial _{\mu }(\theta ^{\beta }*S)  \nonumber \\
&=&\frac{\partial S}{\partial \theta ^{\alpha }}+i\gamma _{\alpha \beta
}^{\mu }\partial _{\mu }\left( \theta ^{\beta }S+\frac{1}{2}C^{\beta \gamma }%
\frac{\partial S}{\partial \theta ^{\gamma }}\right)  \nonumber \\
&\equiv &Q_{\alpha }S,  \label{14}
\end{eqnarray}
This amounts to defining the deformed generators $Q_{\alpha }$ as 
\begin{equation}
Q_{\alpha }\equiv {\cal Q}_{\alpha }+\frac{i}{2}\gamma _{\alpha \beta }^{\mu
}C^{\beta \gamma }\frac{\partial }{\partial \theta ^{\gamma }}\partial _{\mu
},  \label{15}
\end{equation}
or in components 
\begin{eqnarray}
Q_{+} &\equiv &{\cal Q}_{+}+\frac{i}{2}\left( C^{++}\frac{\partial }{%
\partial \theta ^{+}}+C^{+-}\frac{\partial }{\partial \theta ^{-}}\right)
\partial _{+},  \nonumber \\
Q_{-} &\equiv &{\cal Q}_{-}+\frac{i}{2}\left( C^{-+}\frac{\partial }{%
\partial \theta ^{+}}+C^{--}\frac{\partial }{\partial \theta ^{-}}\right)
\partial _{-}.  \label{16}
\end{eqnarray}
The algebra obeyed by the deformed generators simply follows from these
expressions to be 
\begin{eqnarray}
Q_{+}^{2} &=&i\partial _{+}-\frac{1}{2}C^{++}\partial _{+}^{2},  \nonumber \\
Q_{-}^{2} &=&i\partial _{-}-\frac{1}{2}C^{--}\partial _{-}^{2},  \nonumber \\
\{Q_{+},Q_{-}\} &=&-C^{+-}\partial _{+}\partial _{-}.  \label{17}
\end{eqnarray}
This can be thought as a closed algebra if one considers $\partial
_{+}^{2},\partial _{-}^{2},\partial _{+}\partial _{-}$ on its r.h.s. as the
``new'' operators, which commute with $Q_{+},Q_{-}$. The deformed
supertranslation algebra (\ref{17}) is in contrast to another one proposed
in \cite{15,16} (see also \cite{17}) via introducing a gauge connection on
superspace.

In a similar fashion we can define the deformed supercovariant derivatives
based on the corresponding undeformed derivatives, namely 
\begin{eqnarray}
D_{\alpha } &\equiv &{\cal D}_{\alpha }-\frac{i}{2}\gamma _{\alpha \beta
}^{\mu }C^{\beta \gamma }\frac{\partial }{\partial \theta ^{\gamma }}%
\partial _{\mu },  \nonumber \\
{\cal D}_{\alpha } &\equiv &\frac{\partial }{\partial \theta ^{\alpha }}%
-i\gamma _{\alpha \beta }^{\mu }\theta ^{\beta }\partial _{\mu }.  \label{18}
\end{eqnarray}
One can verify that (unlike in the undeformed case as well as in the
deformed 4d case \cite{3}) these have nonvanishing anticommutators with the
deformed SUSY generators, 
\begin{eqnarray}
\{D_{+},Q_{+}\} &=&C^{++}\partial _{+}^{2},\qquad
\{D_{+},Q_{-}\}=C^{+-}\partial _{+}\partial _{-},  \nonumber \\
\{D_{+},Q_{+}\} &=&C^{++}\partial _{+}^{2},\qquad
\{D_{-},Q_{+}\}=C^{-+}\partial _{-}\partial _{+}.  \label{19}
\end{eqnarray}
So, after all, any superspace lagrangian constructed on the basis of the
deformed superderivatives will not be invariant under the deformed SUSY
transformations. Using these derivatives, however, the most natural
definition for the deformed version of the kinetic term of a real scalar
superfield $S$ is written as follows 
\begin{equation}
2{\cal L}_{K}[S]=D_{+}S*D_{-}S\cong D_{+}SD_{-}S.  \label{20}
\end{equation}
Unlike in ordinary noncommutative spaces, here the deformed kinetic term
still contains terms depending on $C$, even after separating its total
derivative part. To extract the $C$-dependence of this expression, we use
the definition (\ref{18}) of $D_{\alpha }$ together with the following
identities 
\begin{eqnarray}
\frac{\partial f}{\partial \theta ^{\pm }}\frac{\partial g}{\partial \theta
^{\pm }} &=&\frac{\partial }{\partial \theta ^{\pm }}\left( f\frac{\partial g%
}{\partial \theta ^{\pm }}\right) \cong 0,  \nonumber \\
\frac{\partial f}{\partial \theta ^{+}}\frac{\partial g}{\partial \theta ^{-}%
} &\cong &-\frac{\partial f}{\partial \theta ^{-}}\frac{\partial g}{\partial
\theta ^{+}},  \label{21}
\end{eqnarray}
from which we find 
\begin{eqnarray}
2{\cal L}_{K}[S] &\cong &{\cal D}_{+}S{\cal D}_{-}S+\frac{i}{2}C^{++}{\cal D}%
_{-}S\partial _{+}\frac{\partial S}{\partial \theta ^{+}}-\frac{i}{2}C^{--}%
{\cal D}_{+}S\partial _{-}\frac{\partial S}{\partial \theta ^{-}}  \nonumber
\\
&&+\frac{1}{2}C^{+-}\left( \theta ^{+}\partial _{+}\frac{\partial S}{%
\partial \theta ^{+}}\partial _{-}S-\theta ^{-}\partial _{-}\frac{\partial S%
}{\partial \theta ^{-}}\partial _{+}S\right)  \nonumber \\
&&-\frac{1}{4}\det C\,\partial _{+}\frac{\partial S}{\partial \theta ^{+}}%
\partial _{-}\frac{\partial S}{\partial \theta ^{-}}.  \label{22}
\end{eqnarray}
Surprisingly, we find that all the $F$-term contributions (i.e. $\theta ^{2}$
components) to this lagrangian which are linear in $C^{\alpha \beta }$ are
in the form of total derivatives and so we are left with the ordinary
lagrangian plus a term proportional to $\det C$. More explicitly, by putting 
\begin{equation}
S=\phi (x)+\bar{\theta}\psi (x)+\frac{1}{2}\bar{\theta}\theta F(x)
\label{23}
\end{equation}
in the above expression, we find for its several terms 
\[
({\cal D}_{+}S{\cal D}_{-}S)_{\theta ^{2}}=-\partial _{+}\phi \partial
_{-}\phi +i\psi _{+}\partial _{-}\psi _{+}+i\psi _{-}\partial _{+}\psi
_{-}+F^{2} 
\]
\begin{eqnarray*}
\left( {\cal D}_{-}S\partial _{+}\frac{\partial S}{\partial \theta ^{+}}%
\right) _{\theta ^{2}} &=&i\partial _{-}\psi _{+}\partial _{+}\psi
_{+}-F\partial _{+}F \\
&=&\frac{i}{2}\left[ \partial _{-}(\psi _{+}\partial _{+}\psi _{+})-\partial
_{+}(\psi _{+}\partial _{-}\psi _{+})\right] -\frac{1}{2}\partial _{+}F^{2},
\end{eqnarray*}
\begin{eqnarray*}
\left( {\cal D}_{+}S\partial _{-}\frac{\partial S}{\partial \theta ^{-}}%
\right) _{\theta ^{2}} &=&i\partial _{+}\psi _{-}\partial _{-}\psi
_{-}-F\partial _{-}F \\
&=&\frac{i}{2}\left[ \partial _{+}(\psi _{-}\partial _{-}\psi _{-})-\partial
_{-}(\psi _{-}\partial _{+}\psi _{-})\right] -\frac{1}{2}\partial _{-}F^{2},
\end{eqnarray*}
\begin{eqnarray*}
&&\left( \theta ^{+}\partial _{+}\frac{\partial S}{\partial \theta ^{+}}%
\partial _{-}S\right) _{\theta ^{2}}-\left( \theta ^{-}\partial _{-}\frac{%
\partial S}{\partial \theta ^{-}}\partial _{+}S\right) _{\theta ^{2}} \\
&=&(-\partial _{+}\psi _{+}\partial _{-}\psi _{-}+\partial _{+}F\partial
_{-}\phi )-(-\partial _{+}\psi _{+}\partial _{-}\psi _{-}+\partial _{+}\phi
\partial _{-}F) \\
&=&\partial _{+}(F\partial _{-}\phi )-\partial _{-}(F\partial _{+}\phi ),
\end{eqnarray*}
\begin{equation}
\left( \partial _{+}\frac{\partial S}{\partial \theta ^{+}}\partial _{-}%
\frac{\partial S}{\partial \theta ^{-}}\right) _{\theta ^{2}}=\partial
_{+}F\partial _{-}F.  \label{24}
\end{equation}
Ignoring the surface terms, the final expression for the ordinary ($x$%
-space) ${\cal L}_{K}$ thus becomes 
\begin{equation}
{\cal L}_{K}=\frac{1}{2}\left( \partial _{+}\phi \partial _{-}\phi +\frac{1}{%
4}\det C\,\partial _{+}F\partial _{-}F-i\psi _{+}\partial _{-}\psi
_{+}-i\psi _{-}\partial _{+}\psi _{-}-F^{2}\right) .  \label{25}
\end{equation}
We observe that, unlike in the commutative theory, the auxiliary field $F$
becomes a dynamical field which can not be solved in terms of the other
variables algebraically. This is also in contrast to the role of the
auxiliary field in SUSY field theories on the ordinary noncommutative spaces
where there also this field appears through its derivatives, though to
infinite order of its derivatives \cite{13b}. We note that the kinetic term
of $F$ will have the right sign (the same as for $\phi $), provided we
choose $\det C>0$.

\subsection{The Model}

We choose as our model the deformed WZ action of a single scalar superfield
which in general consists of the above deformed kinetic lagrangian of $S$
plus a deformed scalar superpotential $W_{*}(S)$. Though, for our purpose in
this paper, there is no essential restriction on the form of the undeformed
superpotential $W(S)$, it is most convenient to work with the cubic
superpotential 
\begin{equation}
W(S)=\frac{1}{2}mS^{2}+\frac{1}{3}gS^{3}.  \label{26}
\end{equation}
By the rules of the previous section, the deformed superpotential then takes
the form 
\begin{eqnarray}
W_{*}(S) &=&\frac{1}{2}mS_{*}^{2}+\frac{1}{3}gS_{*}^{3}  \nonumber \\
&\cong &\frac{1}{2}mS^{2}+\frac{1}{3}g\left[ S^{3}-\frac{1}{4}\det
C\,S\left( \frac{\partial ^{2}S}{\partial \theta ^{2}}\right) ^{2}\right] 
\nonumber \\
&=&W(S)-\frac{1}{12}g\det C\,S\left( \frac{\partial ^{2}S}{\partial \theta
^{2}}\right) ^{2}.  \label{27}
\end{eqnarray}
This contributes to the $F$-term as 
\begin{equation}
\lbrack W_{*}(S)]_{\theta ^{2}}=FW^{\prime }(\phi )-\psi _{+}\psi
_{-}W^{\prime \prime }(\phi )-\frac{1}{12}g\det C\,F^{3}.  \label{28}
\end{equation}
As such, the overall lagrangian of this system, 
\begin{equation}
{\cal L}=-\int d^{2}\theta \left( \frac{1}{2}D_{+}S*D_{-}S+W_{*}(S)\right) ,
\label{29}
\end{equation}
takes the explicit form 
\begin{eqnarray}
{\cal L} &=&\frac{1}{2}\partial _{+}\phi \partial _{-}\phi +\frac{1}{8}\det
C\,\partial _{+}F\partial _{-}F-\frac{i}{2}\bar{\psi}\partial \kern %
-.5em/\psi  \nonumber \\
&&+\frac{1}{12}g\det C\,F^{3}-\frac{1}{2}F^{2}-FW^{\prime }(\phi )+\frac{1}{2%
}\bar{\psi}\psi W^{\prime \prime }(\phi ).  \label{30}
\end{eqnarray}
In later sections, we will consider the $\psi =0$ configurations of this
theory which provide trivial solutions to the EOM of fermions, and will
focus on analyzing its solitons.

\section{Solitons}

The main question we are concerned in this paper is to find the finite
energy solutions of the above model which are continuously connected to the
BPS solutions of the corresponding undeformed model. The bosonic part of
this model, obtained by putting $\psi =0$ in eq.(\ref{30}), reads 
\begin{equation}
{\cal L}[\phi ,F]=\frac{1}{2}\partial _{+}\phi \partial _{-}\phi +\frac{%
\lambda }{8}\partial _{+}F\partial _{-}F+V(\phi ,F),  \label{30a}
\end{equation}
with a potential term of the form 
\begin{equation}
V(\phi ,F)=\frac{\lambda }{12}gF^{3}-\frac{1}{2}F^{2}-FW^{\prime }(\phi ),
\label{30b}
\end{equation}
where $W^{\prime }(\phi )=m\phi +g\phi ^{2}$ and for convenience we have
denoted the deformation parameter by $\lambda \equiv \det C.$ This is an
example of coupled scalar field theories in 2 dimensions which finding their
exact solitonic solutions is not a generally well posed problem (see,
however, the methods in \cite{30}). It becomes, however easy to determine
such a solution, if an ``orbit'' equation in the $(\phi ,F)$ plane,
connecting two specific ``boundary points'', has been specified via some
other information. The boundary points must themselves be solutions to the
EOM and hence they satisfy 
\begin{eqnarray}
\frac{\partial V}{\partial \phi } &=&-FW^{\prime \prime }(\phi )=0, 
\nonumber \\
\frac{\partial V}{\partial F} &=&\frac{\lambda }{4}gF^{2}-F-W^{\prime }(\phi
)=0.  \label{33c}
\end{eqnarray}
A solution of these equations not depending on $\lambda $ is given by $%
F=W^{\prime }(\phi )=0$ (i.e. $\phi =0$ or $-m/g$), which specifies the
classical supersymmetric vacua of the undeformed theory. We choose to work
with these boundary values of $(\phi ,F)$ for the rest of this section.

Though we are not claiming to obtain all the solutions in this paper, in the
following we will find via perturbative method a class of solutions which
are continuous deformations of the standard BPS solutions. For this, let us
consider the time independent solutions of this system which are governed by
the following 1-dimensional lagrangian,

\begin{equation}
{\cal L}[\phi ,F]=\frac{1}{2}\left( \frac{d\phi }{dx}\right) ^{2}+\frac{%
\lambda }{8}\left( \frac{dF}{dx}\right) ^{2}+V(\phi ,F),  \label{31}
\end{equation}
describing motion of an analogue particle in 2 dimensions (with $x$ as the
time variable).$\,$\footnote{%
Here we adapt the convention $x^{\pm }=\frac{1}{2}(x\pm iy),$ with $x$ the
spacial coordinate and $y$ the eucilean time.} The EOM of $\phi ,F$ from
this lagrangian follow to be 
\begin{eqnarray}
\frac{d^{2}\phi }{dx^{2}} &=&-FW^{\prime \prime }(\phi ),  \label{33a} \\
\frac{\lambda }{4}\frac{d^{2}F}{dx^{2}} &=&\frac{\lambda }{4}%
gF^{2}-F-W^{\prime }(\phi ).  \label{33b}
\end{eqnarray}
Useful information regarding this system of equations can be obtained from a
conserved quantity corresponding to the ``energy'' of the analogue particle
described by eq.(\ref{31}). This has the form 
\begin{equation}
{\cal E}=\frac{1}{2}\left( \frac{d\phi }{dx}\right) ^{2}+\frac{\lambda }{8}%
\left( \frac{dF}{dx}\right) ^{2}-V(\phi ,F).  \label{34}
\end{equation}
The finite energy solutions of the original theory are specified as those
solutions of the system (\ref{31}) for which the energy of the analogue
particle vanishes, i.e. ${\cal E}=0$ \cite{30}. Hence solutions of interest
should satisfy the ``conservation equation'': 
\begin{equation}
\frac{1}{2}\left( \frac{d\phi }{dx}\right) ^{2}+\frac{\lambda }{8}\left( 
\frac{dF}{dx}\right) ^{2}=V(\phi ,F).  \label{35}
\end{equation}

On the other hand, we know that for the undeformed theory with $\lambda =0$,
the finite energy BPS solutions are given by the following equations \cite
{29} 
\begin{eqnarray}
F &=&-W^{\prime }(\phi ),  \nonumber \\
\frac{d\phi }{dx} &=&\pm W^{\prime }(\phi ).  \label{36}
\end{eqnarray}
As we mentioned, here we are interested in those solutions of (\ref{31})
that are continuously connected to the solutions of eqs.(\ref{36}), when $%
\lambda $ slightly differs from zero. Such solutions will have series
expansions in powers of $\lambda $ of the form 
\begin{eqnarray}
\phi (x) &=&\phi _{0}(x)+\lambda \phi _{1}(x)+\lambda ^{2}\phi _{2}(x)+\cdot
\cdot \cdot ,  \nonumber \\
F(x) &=&F_{0}(x)+\lambda F_{1}(x)+\lambda ^{2}F_{2}(x)+\cdot \cdot \cdot ,
\label{37}
\end{eqnarray}
with $\phi _{0}(x),F_{0}(x)$ denoting a solution of the unperturbed eqs.(\ref
{36}). To find a perturbative scheme for determining such solutions, we
first write eq.(\ref{33b}). as 
\begin{equation}
F=-W^{\prime }(\phi )-\frac{\lambda }{4}\left( \frac{d^{2}F}{dx^{2}}%
-gF^{2}\right) .  \label{38}
\end{equation}
Here we limit our discussion only to first order corrections, though the
procedure can indeed be extended to all orders in $\lambda $. The above
equation suggests an iterative scheme by which we can first determine
perturbatively the ``orbit'' equation, $F=F(\phi )$, and subsequently find
the expression for $d\phi /dx$ in terms of $\phi $ at each order of $\lambda 
$. We start iteration by putting $F=-W^{\prime }(\phi )+{\cal O}(\lambda ) $
on the r.h.s. of eq. (\ref{38}), from which follows 
\begin{eqnarray}
F &=&-W^{\prime }(\phi )-\frac{\lambda }{4}\left( -\frac{d^{2}W^{\prime
}(\phi )}{dx^{2}}-gW^{\prime 2}(\phi )+{\cal O}(\lambda )\right)  \nonumber
\\
&=&-W^{\prime }-\frac{\lambda }{4}\left[ -W^{\prime \prime }\frac{d^{2}\phi 
}{dx^{2}}-W^{(3)}\left( \frac{d\phi }{dx}\right) ^{2}-gW^{\prime 2}\right] +%
{\cal O}(\lambda ^{2}).  \label{39}
\end{eqnarray}
Now, using the EOM\ of $\phi $, eq.(\ref{34}), and that $W^{(3)}=2g$, the
above equation becomes 
\begin{eqnarray}
F &=&-W^{\prime }-\frac{\lambda }{4}\left[ FW^{\prime \prime 2}-2g\left( 
\frac{d\phi }{dx}\right) ^{2}-gW^{\prime 2}\right] +{\cal O}(\lambda ^{2}) 
\nonumber \\
&=&-W^{\prime }+\frac{\lambda }{4}\left[ W^{\prime }W^{\prime \prime
2}+2g\left( \frac{d\phi }{dx}\right) ^{2}+gW^{\prime 2}\right] +{\cal O}%
(\lambda ^{2}),  \label{40}
\end{eqnarray}
where in the last step we have used of $F=-W^{\prime }(\phi )+{\cal O}%
(\lambda )$ once again. The second term in the brackets equals $2gW^{\prime
2}$ to order ${\cal O}(\lambda )$, as by assumption the BPS equation in our
case deviates to $\frac{d\phi }{dx}=\pm W^{\prime }(\phi )+{\cal O}(\lambda
).$\footnote{%
We can show this simply by putting the zeroth order equation $F=-W^{\prime
}(\phi )+{\cal O}(\lambda )$ into the EOM of $\phi $ giving rise to $\frac{%
d^{2}\phi }{dx^{2}}=W^{\prime }(\phi )W^{\prime \prime }(\phi )+$ ${\cal O}%
(\lambda )$ and integrating it to $\left( \frac{d\phi }{dx}\right)
^{2}=W^{\prime 2}(\phi )+$ ${\cal O}(\lambda )$ using the boundary condition
that $\phi (x)\rightarrow $const. at $x\rightarrow \pm \infty .$} Thus the
desired equation of orbit $F=F(\phi )$ to first order in $\lambda $ becomes 
\begin{equation}
F=-W^{\prime }(\phi )+\frac{\lambda }{4}W^{\prime }(\phi )\left( W^{\prime
\prime 2}(\phi )+3gW^{\prime }(\phi )\right) +\cdot \cdot \cdot .  \label{41}
\end{equation}
(Hereafter by ellipsis we mean the terms which are higher than first order
in $\lambda $.) We note that to this order of $\lambda $, $\;F(\phi )$ is a
polynomial function of $\phi $ which is divisible by the polynomial $%
W^{\prime }(\phi )$. This is indeed a generic property of $F(\phi )$
repeated to all orders in $\lambda $. Since the roots of $W^{\prime }(\phi )$
specify the SUSY vacua of the ordinary theory, after all the above
expression shows that the $F=W^{\prime }(\phi )=0$ solutions (like in the
ordinary theory) still describe those vacua of the deformed theory which are
invariant under the ordinary SUSY.

We now determine an expression for $d\phi /dx$ in terms of $\phi $ to order $%
{\cal O}(\lambda )$, from which the first order correction to $\phi _{0}(x)$
is obtained. One way to do this is to apply the equation of orbit (eq.(\ref
{41}) ) directly into the EOM\ of $\phi $ ( eq.(\ref{33a}) ). This gives 
\begin{equation}
\frac{d^{2}\phi }{dx^{2}}=W^{\prime }(\phi )W^{\prime \prime }(\phi )-\frac{%
\lambda }{4}W^{\prime }(\phi )W^{\prime \prime }(\phi )\left( W^{\prime
\prime 2}(\phi )+3gW^{\prime }(\phi )\right) +\cdot \cdot \cdot .  \label{42}
\end{equation}
Upon multiplication by $2d\phi /dx$ and subsequent integration this leads to 
\begin{equation}
\left( \frac{d\phi }{dx}\right) ^{2}=W^{\prime 2}(\phi )-\frac{\lambda }{2}%
\int d\phi \left( W^{\prime }(\phi )W^{\prime \prime 3}(\phi )+3gW^{\prime
2}(\phi )W^{\prime \prime }(\phi )\right) +\cdot \cdot \cdot ,  \label{43}
\end{equation}
where the second term on the r.h.s. is an indefinite integral. Although, the
integration here is straightforward due to the polynomial nature of $W(\phi
) $, it is helpful to express it in a more explicit form. For this, we use
by part integration to write 
\begin{eqnarray}
&&\int d\phi \left( W^{\prime }W^{\prime \prime 3}+3gW^{\prime 2}W^{\prime
\prime }\right)  \nonumber \\
&=&\int d\phi \frac{1}{2}(W^{\prime 2})^{^{\prime }}W^{\prime \prime
2}+gW^{\prime 3}  \nonumber \\
&=&\frac{1}{2}\left( W^{\prime 2}W^{\prime \prime 2}-\int d\phi W^{\prime
2}(2W^{\prime \prime }W^{(3)})\right) +gW^{\prime 3}  \nonumber \\
&=&\frac{1}{2}W^{\prime 2}W^{\prime \prime 2}+\frac{g}{3}W^{\prime 3},
\label{44}
\end{eqnarray}
where we have used the fact that $W^{(3)}=2g$ is constant and have adjusted
the integration constant such that the overall result vanishes at the
critical points of $W(\phi )$.

We can formulate the above procedure for determining $d\phi /dx$ to
arbitrary order of $\lambda $ in a more systematic way by forming the
effective lagrangian of the single variable $\phi (x)$, given that the
desired solution obeys the known orbit equation $F=F(\phi )$. This is most
easily achieved by replacing $F$ in terms of $\phi $ in the lagrangian (\ref
{31}), upon which it becomes 
\begin{eqnarray}
{\cal L}_{{\rm {eff}}}[\phi ] &=&\frac{1}{2}\left[ 1+\frac{\lambda }{4}%
F^{\prime 2}(\phi )\right] \left( \frac{d\phi }{dx}\right) ^{2}+V(\phi
,F(\phi ))  \nonumber \\
&\equiv &\frac{1}{2}G(\phi )\left( \frac{d\phi }{dx}\right) ^{2}+U(\phi ).
\label{45}
\end{eqnarray}
Here we defined the effective ``metric'' $G(\phi )$ and the effective scalar
potential $U(\phi )$ by the quantities in the first line. Using the known
expression for $F(\phi )$ it is easy to work out explicit expressions for $%
G(\phi ),U(\phi )$ to each order of $\lambda $. For example, to first order
we find the expressions 
\begin{eqnarray*}
G(\phi ) &\equiv &1+\frac{\lambda }{4}F^{\prime 2}(\phi ) \\
&=&1+\frac{\lambda }{4}W^{\prime \prime 2}(\phi )+\cdot \cdot \cdot ,
\end{eqnarray*}
\begin{eqnarray}
U(\phi ) &\equiv &\frac{\lambda }{12}gF^{3}(\phi )-\frac{1}{2}F^{2}(\phi
)-F(\phi )W^{\prime }(\phi )  \nonumber \\
&=&\frac{1}{2}W^{\prime 2}(\phi )-\frac{\lambda }{12}gW^{\prime 3}(\phi
)+\cdot \cdot \cdot .  \label{46}
\end{eqnarray}
Evidently, for $\lambda \geq 0$ (as we assumed earlier), we have a positive
definite metric $G(\phi )>0$. Moreover, we see that (at least to this order
of $\lambda $) the modification to the standard effective potential $U(\phi
) $ $=\frac{1}{2}W^{\prime 2}(\phi )$ is proportional to $W^{\prime 3}(\phi
) $, and so both $U(\phi )$ and its derivative $U^{\prime }(\phi )$ vanish
at all the critical points of the superpotential $W(\phi )$, so that $U(\phi
)$ is always positive in a small vicinity of such points. However, we note
that (unlike in the ordinary case), $U(\phi )$ is {\it not} a positive
definite function of $\phi $ and it so may possess additional zeros given by
the roots of $1-\frac{\lambda }{6}gW^{\prime }(\phi )+\cdot \cdot \cdot $.
These would correspond to a class of solutions of the deformed theory which
are not continuously connected to a solution of the undeformed theory via a
power series expansion in $\lambda $. The critical point of $U(\phi )$ (i.e.
the common roots of $U(\phi ),U^{\prime }(\phi )$ ) specify the boundary
values of $\phi (x)$ at $x\rightarrow \pm \infty $ in order for the soliton
to have a finite energy.

Here we are interested in those solutions of the deformed theory that have
the same boundary conditions as those of the undeformed theory. Therefore we
assume $\phi (x)\rightarrow \phi _{\pm \infty }$ at $x\rightarrow \pm \infty 
$, so that 
\begin{equation}
W^{\prime }(\phi _{\infty })=W^{\prime }(\phi _{-\infty })=0.  \label{47}
\end{equation}
The soliton equation as found from the ${\cal E}=0$ condition now becomes 
\begin{equation}
{\cal E}=\frac{1}{2}G(\phi )\left( \frac{d\phi }{dx}\right) ^{2}-U(\phi )=0,
\label{48}
\end{equation}
which upon using the ${\cal O}(\lambda )$ expressions of $G(\phi )$ and $%
U(\phi )$ leads to 
\begin{equation}
\pm \frac{d\phi }{dx}=\sqrt{\frac{2U(\phi )}{G(\phi )}}=W^{\prime }(\phi )-%
\frac{\lambda }{4}\left( \frac{g}{3}W^{\prime 2}(\phi )+\frac{1}{2}W^{\prime
}(\phi )W^{\prime \prime 2}(\phi )\right) +\cdot \cdot \cdot .  \label{49}
\end{equation}
This is just the same result we arrived in eqs.(\ref{43}),(\ref{44})
directly using the EOM.

We can solve the differential equation (\ref{49}) perturbatively for $\phi
(x)=\phi _{0}(x)+\lambda \phi _{1}(x)+\cdot \cdot \cdot $ to arbitrary order
in $\lambda $. For this, let us write this equation simply as 
\begin{equation}
\pm \frac{d\phi }{dx}=v_{0}(\phi )+\lambda v_{1}(\phi )+\cdot \cdot \cdot ,
\label{50}
\end{equation}
where 
\begin{eqnarray}
v_{0}(\phi ) &\equiv &W^{\prime }(\phi ),  \nonumber \\
v_{1}(\phi ) &\equiv &-\frac{1}{4}W^{\prime }(\phi )\left( \frac{g}{3}%
W^{\prime }(\phi )+\frac{1}{2}W^{\prime \prime 2}(\phi )\right) ,  \label{51}
\end{eqnarray}
and so on. The zeroth order (unperturbed) solution $\phi _{0}(x)$ then
satisfies the unperturbed BPS equation 
\begin{equation}
\pm \frac{d\phi _{0}}{dx}=v_{0}(\phi _{0}),  \label{52}
\end{equation}
having the standard kink/anti-kink solution \cite{30} 
\begin{equation}
\phi _{0}(x)=\frac{m}{2g}\left( \pm \tanh \frac{mx}{2}-1\right) .  \label{53}
\end{equation}

\begin{figure}[tbh]
\centerline{\includegraphics[width=0.51\textwidth]{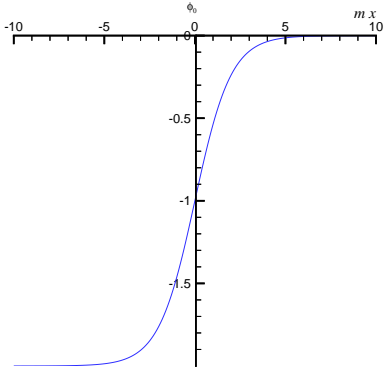}}
\caption{{{{{{{\protect\small {The profile of the undeformed kink soliton, ${%
\phi}_0(x)$, with ${\phi}_0$ measured in units of $\frac{m}{2g}$.}}}}}}}}
\end{figure}

The equation for first order correction $\phi _{1}(x)$ on the other hand
becomes 
\begin{equation}
\pm \frac{d\phi _{1}}{dx}-v_{0}^{\prime }(\phi _{0})\phi _{1}=v_{1}(\phi
_{0}).  \label{54}
\end{equation}
This is a linear ODE of first order with non-constant coefficients whose
solution for $\phi _{1}(x)$ is known in terms of quadratures: 
\begin{equation}
\phi _{1}=\pm e^{\pm \int dxv^{\prime }(\phi _{0})}\int dxv_{1}(\phi
_{0})e^{\mp \int dxv^{\prime }(\phi _{0})}.  \label{55}
\end{equation}
This solution can be simplified by using $\pm \frac{d\phi _{0}}{dx}%
=v_{0}(\phi _{0})$ from which we obtain 
\begin{equation}
\pm \int dxv_{0}^{\prime }(\phi _{0})=\int d\phi _{0}\frac{v_{0}^{\prime
}(\phi _{0})}{v_{0}(\phi _{0})}=\log v_{o}(\phi _{0}),  \label{56}
\end{equation}
and therefore for $\phi _{1}$ in terms of $\phi _{0}$%
\begin{equation}
\phi _{1}=\pm v_{0}(\phi _{0})\int dx\frac{v_{1}(\phi _{0})}{v_{0}(\phi _{0})%
}=v_{0}(\phi _{0})\int d\phi _{0}\frac{v_{1}(\phi _{0})}{v_{0}^{2}(\phi _{0})%
}.  \label{57}
\end{equation}
Using the explicit expression for $v_{0}(\phi ),v_{1}(\phi )$ in terms of $%
W(\phi )$ this becomes 
\begin{eqnarray}
\phi _{1} &=&-\frac{1}{4}W^{\prime }(\phi _{0})\int \frac{d\phi _{0}}{%
W^{\prime }(\phi _{0})}\left( \frac{g}{3}W^{\prime }(\phi _{0})+\frac{1}{2}%
W^{\prime \prime 2}(\phi _{0})\right)  \nonumber \\
&=&(m\phi _{0}+g\phi _{0}^{2})\left[ -\frac{7}{12}g\phi _{0}+\frac{m}{8}\log
\left( \frac{m+g\phi _{0}}{g\phi _{0}}\right) +c\right] .  \label{58}
\end{eqnarray}

\begin{figure}[tbh]
\centerline{\includegraphics[width=0.51\textwidth]{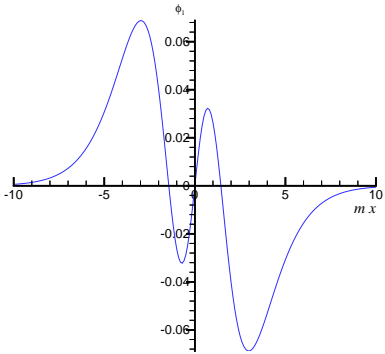}}
\caption{{{{{{{\protect\small {The profile of the ${\cal O}(\lambda )$
correction to the kink soliton, ${\phi}_1(x)$, with ${\phi}_1$ measured in
units of $\frac{7m^3}{96g}$.}}}}}}}}
\end{figure}

Finally, using the known solution for $\phi _{0}(x),$ this solution is
translated into an explicit expression for $\phi _{1}(x)$%
\begin{equation}
\phi _{1}(x)=\pm \frac{7m^{3}}{96g}\left( \frac{\tanh \frac{mx}{2}-\frac{3}{7%
}mx}{\cosh ^{2}\frac{mx}{2}}\right) ,  \label{59}
\end{equation}
where the integration constant is chosen $c=0$ so that at $x=0$ the
correction does not change the value of $\phi (0)=-\frac{m}{2g}$, i.e. so
that $\phi _{1}(0)=0$. Note that the first order correction to either of the
kink/anti-kink solutions is an antisymmetric function of $x$ and it vanishes
at $x\rightarrow \pm \infty ,$ as we have expected. The profiles of $\phi
_{0}(x),\phi _{1}(x)$ are plotted in figs.1,2 below.

\section{The Effective Superpotential and Modification of the BPS Mass
Formula}

As we mentioned in the last section, solitons of interest in the deformed
theory are solutions of an effective 1-dimensional system described by the
lagrangian (\ref{45}). This lagrangian can be imagined to have originated
from a non-linear $\sigma $-model on ${\cal N}=1,D=2$ superspace with the
lagrangian:\footnote{%
Here $\hat{S}$ and $S$ have the same $\phi $ and $\psi $ components$,$ but
their auxiliary components $\hat{F}$ and $F$ are different, and indeed,
unlike the latter the former is a non-dynamical field which on its EOM\ is
algebraically solved in terms of $\phi $ as $\hat{F}=-{\cal W}^{\prime
}(\phi )/G(\phi )$.} 
\begin{equation}
{\cal L}_{{\rm {eff}}}[\hat{S}]=\frac{1}{2}G(\hat{S}){\cal D}_{+}S{\cal D}%
_{-}S+{\cal W}(\hat{S}).  \label{60}
\end{equation}
This is a special (one-dimensional) case of the more general $\sigma $-model
with multi-dimensional target space analyzed in an appendix in \cite{14}. It
is easy to see that for such a model the field theory scalar potential has
the following form 
\begin{equation}
U(\phi )=\frac{{\cal W}^{\prime 2}(\phi )}{2G(\phi )}.  \label{61}
\end{equation}
Conversely, we can solve for ${\cal W}(\phi )$ in terms of the known
functions $G(\phi )$, $U(\phi )$ as follows 
\begin{eqnarray}
{\cal W}(\phi ) &=&\int d\phi [2G(\phi )U(\phi )]^{1/2}  \nonumber \\
&=&\int d\phi \left[ \left( 1+\frac{\lambda }{4}W^{\prime \prime 2}+\cdot
\cdot \cdot \right) \left( W^{\prime 2}-\frac{\lambda }{6}gW^{\prime
3}+\cdot \cdot \cdot \right) \right] ^{1/2}  \nonumber \\
&=&\int d\phi W^{\prime }(\phi )\left[ 1+\frac{\lambda }{2}\left( \frac{1}{4}%
W^{\prime \prime 2}(\phi )-\frac{g}{6}W^{\prime }(\phi )\right) +\cdot \cdot
\cdot \right] .  \label{62}
\end{eqnarray}
The zeroth order term is simply $W(\phi )$, while the ${\cal O}(\lambda )$
term is simplified using $\int d\phi W^{\prime }W^{\prime \prime 2}=\frac{1}{%
2}W^{\prime 2}W^{\prime \prime }-g\int d\phi W^{\prime 2}$ to give the
result 
\begin{equation}
{\cal W}(\phi )=W(\phi )+\lambda \left( \frac{1}{16}W^{\prime 2}(\phi
)W^{\prime \prime }(\phi )-\frac{5}{24}g\int d\phi W^{\prime 2}(\phi
)\right) +\cdot \cdot \cdot .  \label{63}
\end{equation}

Now for calculation of the mass of the deformed BPS soliton, we recall the
general BPS mass formula \cite{29,14}, which in terms of the effective $%
\sigma $-model (\ref{60}) may be written as 
\begin{equation}
{\cal M}=\left| \int dxd^{2}\theta \,{\cal W}(\hat{S})\right| =\left| {\cal W%
}(\phi _{\infty })-{\cal W}(\phi _{-\infty })\right| .  \label{64}
\end{equation}
Using ${\cal W}(\phi )$ of eq.(\ref{62}) and noting that $W^{\prime }(\phi
_{\pm \infty })=0$, this formula gives ${\cal M}={\cal M}^{(0)}+\lambda 
{\cal M}^{(1)}+\cdot \cdot \cdot ,$where 
\begin{eqnarray}
{\cal M}^{(0)} &=&\left| W(\phi _{\infty })-W(\phi _{-\infty })\right|
=\left| \left( \frac{m}{2}\phi ^{2}+\frac{g}{3}\phi ^{3}\right) _{\phi
=-m/g}^{0}\right| =\frac{m^{3}}{6g^{2}},  \nonumber \\
{\cal M}^{(1)} &=&\frac{5g}{24}\int_{\phi _{-\infty }}^{\phi _{\infty
}}d\phi W^{\prime 2}(\phi )=\frac{5g}{24}\int_{-m/g}^{0}d\phi (m\phi +g\phi
^{2})^{2}=\frac{1}{144}\frac{m^{5}}{g^{2}}.  \label{65}
\end{eqnarray}
In other words 
\begin{equation}
{\cal M}=\frac{m^{3}}{6g^{2}}\left( 1+\frac{1}{24}m^{2}\det C+\cdot \cdot
\cdot \right) .  \label{66}
\end{equation}
As this expression suggests, a meaning of smallness of $C$ concerned to our
computations is $|\det C|\ll 24/m^{2}$. Note that the usual dependence of
the mass of a soliton to the coupling as ${\cal M}\sim 1/g^{2}$ does not get
modified due to the deformation of superspace. In particular, solitons still
have a large/small mass at weak/strong coupling and so they can be
interpreted as non-perturbative objects of field theory.

\section{Non-linear Realization of ${\cal N}=1$ Supersymmetry}

So far we have stressed that the deformed the ${\cal N}=1$ theory in 2d
explicitly breaks all the supersymmetries that underlie the original
undeformed theory. We have noticed, however, that a class of solutions of
the deformed theory can still be described as the $1/2$ BPS solutions of
some other (undeformed) ${\cal N}=1$ supersymmetric $\sigma $-model with a
metric and superpotential deviating from those of the original model by
corrections in powers of the small parameter $\lambda $. The simple fact
that these solutions can be described as the one-half SUSY solutions of an $%
{\cal N}=1$ theory raises the question that whether the deformed ${\cal N}=0$
theory realizes the underlying ${\cal N}=1$ SUSY somehow non-linearly.

To find the answer, let us first write the complete form of the effective $%
{\cal N}=1$ non-linear $\sigma $-model on the orbit of $F=F(\phi )$,
including its fermions, that is 
\begin{eqnarray}
{\cal L}_{{\rm {eff}}}[\phi ,\psi ] &=&\frac{1}{2}G(\phi )\partial _{+}\phi
\partial _{-}\phi +U(\phi )  \nonumber \\
&&-\frac{i}{2}\bar{\psi}\partial \kern -.5em/\psi +\frac{1}{2}W^{\prime
\prime }(\phi )\bar{\psi}\psi ,  \label{67}
\end{eqnarray}
with $G(\phi ),U(\phi )$ given in terms of $W(\phi )$ as in eqs.(\ref{46}).
Actually this ${\cal L}_{{\rm {eff}}}$ is not written in the standard form
of an ${\cal N}=1$ non-linear $\sigma $-model with the auxiliary fields
integrated out using their EOM. To find a precise match let us consider a
generic $\sigma $-model on one-dimensional target space defined by the
action 
\begin{equation}
I[\tilde{S}]=-\int d^{2}xd^{2}\theta \left( \frac{1}{2}\tilde{G}(\tilde{S})%
{\cal D}_{+}\tilde{S}{\cal D}_{-}\tilde{S}+\tilde{W}(\tilde{S})\right) .
\label{68}
\end{equation}
By carrying the $\theta $-integrations and subsequently by eliminating the
auxiliary field $\tilde{F}$ via its EOM, $\tilde{F}=-\tilde{W}(\tilde{\phi})/%
\tilde{G}(\tilde{\phi})$, we find the ordinary space lagrangian 
\begin{eqnarray}
\widetilde{{\cal L}}[\tilde{\phi},\tilde{\psi}] &=&\frac{1}{2}\tilde{G}(%
\tilde{\phi})\partial _{+}\tilde{\phi}\partial _{-}\tilde{\phi}+\frac{\tilde{%
W}^{\prime 2}(\tilde{\phi})}{2\tilde{G}(\tilde{\phi})}  \nonumber \\
&&-\frac{i}{2}\tilde{G}(\tilde{\phi})\overline{\tilde{\psi}}\partial \kern %
-.5em/\tilde{\psi}+\frac{1}{2}\sqrt{\tilde{G}(\tilde{\phi})}\left( \frac{%
\tilde{W}^{\prime }(\tilde{\phi})}{\sqrt{\tilde{G}(\tilde{\phi})}}\right)
^{\prime }\overline{\tilde{\psi}}\tilde{\psi}.  \label{69}
\end{eqnarray}
We now demand that ${\cal L}_{{\rm {eff}}}$ takes the same form as $%
\widetilde{{\cal L}}$ via a suitable map between their field contents. A
simple comparison of similar terms between the two lagrangians then suggests
that this map must be of the form 
\begin{equation}
\tilde{\phi}=\tilde{\phi}(\phi ),\qquad \tilde{\psi}=\frac{1}{\sqrt{\tilde{G}%
(\tilde{\phi})}}\psi .  \label{70}
\end{equation}
The unknown function $\tilde{\phi}(\phi )$ is to be determined along with
the metric $\tilde{G}(\tilde{\phi})$ and superpotential $\tilde{W}(\tilde{%
\phi})$ of the $\sigma $-model. The equivalence between the two lagrangians
then implies the following conditions 
\begin{eqnarray}
\tilde{G}(\tilde{\phi})\left( \frac{d\tilde{\phi}}{d\phi }\right) ^{2}
&=&G(\phi ),  \nonumber \\
\frac{\tilde{W}^{\prime 2}(\tilde{\phi})}{2\tilde{G}(\tilde{\phi})}
&=&U(\phi ),  \nonumber \\
\sqrt{\tilde{G}(\tilde{\phi})}\left( \frac{\tilde{W}^{\prime }(\tilde{\phi})%
}{\sqrt{\tilde{G}(\tilde{\phi})}}\right) ^{\prime } &=&W^{\prime \prime
}(\phi ).  \label{71}
\end{eqnarray}
Here the primes in each case denote differentiations with respect to the
specified variable. These set of equations determine both the map $\tilde{%
\phi}=\tilde{\phi}(\phi )$ and the unknown functions $\tilde{G}(\tilde{\phi}%
) $, $\tilde{W}(\tilde{\phi})$ by quadratures. For this, using the last two
equations in (\ref{71}), we first write 
\begin{equation}
\sqrt{\tilde{G}(\tilde{\phi})}\left( \sqrt{2U(\phi )}\right) ^{\prime }\frac{%
d\phi }{d\tilde{\phi}}=W^{\prime \prime }(\phi ).  \label{72}
\end{equation}
Define for convenience the (known) function of $\phi ,$%
\begin{equation}
H(\phi )\equiv \frac{\left( \sqrt{2U(\phi )}\right) ^{\prime }}{W^{\prime
\prime }(\phi )}.  \label{73}
\end{equation}
Then, combining the previous equation with eqs.(\ref{71}), we obtain 
\begin{eqnarray}
\tilde{G}(\tilde{\phi}(\phi )) &=&\frac{\sqrt{G(\phi )}}{H(\phi )}, 
\nonumber \\
\tilde{W}^{\prime }(\tilde{\phi}(\phi )) &=&\sqrt{\frac{2U(\phi )\sqrt{%
G(\phi )}}{H(\phi )}},  \nonumber \\
\left( \frac{d\tilde{\phi}}{d\phi }\right) ^{4} &=&G(\phi )H^{2}(\phi ).
\label{74}
\end{eqnarray}
These in principle specify the solutions for $\tilde{G}$, $\tilde{W}$ as
functions of $\tilde{\phi},$ provided we have integrated and inverted the
last equation for $\tilde{\phi}(\phi )$ : 
\begin{equation}
\tilde{\phi}=\int d\phi \sqrt[4]{G(\phi )H^{2}(\phi )}.  \label{75}
\end{equation}

Although it is possible to write down the explicit forms of these solutions
perturbatively in powers of $\lambda ,$ it turns out that for determining
the non-linearly deformed SUSY transformations of $\phi $, $\psi $
(corresponding to the ordinary transformations of $\tilde{\phi}$, $\tilde{%
\psi}$) we do not need to know these explicit solutions. To see how this
works, let us begin with recalling the ordinary SUSY transformations of $%
\tilde{\phi}$, $\tilde{\psi}$, i.e. \cite{14} 
\begin{eqnarray}
\delta _{\epsilon }\tilde{\phi} &=&\bar{\epsilon}\tilde{\psi},  \nonumber \\
\delta _{\epsilon }\tilde{\psi} &=&-\left( \partial \kern -.5em/\tilde{\phi}+%
\frac{\tilde{W}^{\prime }(\tilde{\phi})}{\tilde{G}(\tilde{\phi})}-\frac{1}{4}%
\frac{\tilde{G}^{\prime }(\tilde{\phi})}{\tilde{G}(\tilde{\phi})}\overline{%
\tilde{\psi}}\tilde{\psi}\right) \epsilon .  \label{76}
\end{eqnarray}
Putting $(\tilde{\phi},\tilde{\psi})$ in terms of $(\phi ,\psi )$ in these
equations we find 
\begin{eqnarray}
\frac{d\tilde{\phi}}{d\phi }\delta _{\epsilon }\phi &=&\bar{\epsilon}\frac{%
\psi }{\sqrt{\tilde{G}}},  \nonumber \\
\delta _{\epsilon }\left( \frac{\psi }{\sqrt{\tilde{G}}}\right) &=&-\left(
\partial \kern -.5em/\phi \frac{d\tilde{\phi}}{d\phi }+\frac{\tilde{W}%
^{\prime }(\tilde{\phi})}{\tilde{G}(\tilde{\phi})}-\frac{1}{4}\frac{\tilde{G}%
^{\prime }(\tilde{\phi})}{\tilde{G}^{2}(\tilde{\phi})}\bar{\psi}\psi \right)
\epsilon .  \label{77}
\end{eqnarray}
Expanding the variation on the l.h.s. of the second equation above and using
the first equation, together with eqs.(\ref{74}), we arrive at the desired
transformations: 
\begin{eqnarray}
\delta _{\epsilon }\phi &=&\frac{\bar{\epsilon}\psi }{\sqrt{G(\phi )}}, 
\nonumber \\
\delta _{\epsilon }\psi &=&-\left( \sqrt{G(\phi )}\partial \kern -.5em/\phi +%
\sqrt{2U(\phi )}\right) \epsilon ,  \label{78}
\end{eqnarray}
which leads to known expressions without really requiring to solve eqs.(\ref
{71}). These must be of course accompanied by the transformation equation of
the auxiliary field $F$, which using the orbit equation $F=F(\phi )$ is
found to be 
\begin{equation}
\delta _{\epsilon }F=\frac{F^{\prime }(\phi )}{\sqrt{G(\phi )}}\bar{\epsilon}%
\psi .  \label{79}
\end{equation}
These deformed transformations can explicitly be expanded in powers of $%
\lambda $ and can be checked that they are actually symmetries of the
lagrangian (\ref{67}) to arbitrary order in $\lambda $. Also, one can check
that the equation $\delta _{\epsilon }\psi =0$ (the equation $\delta
_{\epsilon }\phi =0$ becomes trivial for $\psi =0$) leads to the same eq.(%
\ref{49}) which we derived previously for the deformation of the BPS
solutions.

\end{document}